# Enhancing kidney quality assessment: Power Doppler during normothermic machine perfusion

Yitian Fang[1*], Anton V. Nikolaev[2*], Jeroen Essers[3, 4, 5], Gisela Ambagtsheer[1], Marian C. Clahsen-van Groningen[6, 7], Robert C. Minnee[1], Ron W.F. de Bruin[1], Gijs van Soest[2, 8]

1. Division of HPB and Transplant Surgery, Department of Surgery, Transplant Institute, Erasmus Medical Center, Rotterdam, the Netherlands.
2. Cardiovascular Institute, Thorax Center, Department of Cardiology, Erasmus Medical Center, Rotterdam, the Netherlands.
3. Department of Molecular Genetics, Cancer Genomics Center, Erasmus Medical Center, Rotterdam, the Netherlands.
4. Department of Radiotherapy, Erasmus Medical Center, Rotterdam, the Netherlands.
5. Department of Vascular Surgery, Cardiovascular Institute, Erasmus Medical Center, Rotterdam, the Netherlands.
6. Department of Pathology and Clinical Bioinformatics, Erasmus Medical Center, Rotterdam, the Netherlands.
7. Department of Medicine 2 (Medical Faculty), RWTH Aachen University, Aachen, Germany.
8. Department of Precision and Microsystems Engineering, Faculty of Mechanical Engineering, Delft University of Technology, Delft, the Netherlands.

* Authors contributed equally to this study.

**Correspondence**

Gijs van Soest

Cardiovascular Institute, Thorax Center, Department of Cardiology, Erasmus Medical Center.

Dr. Molewaterplein 40, 3015 GD Rotterdam, The Netherlands.

Email: g.vansoest@erasmusmc.nl

**Abstract**

**Objectives** Marginal donor kidneys are increasingly used for transplantation to overcome organ shortage. This study aims to investigate the additional value of Power Doppler (PD) imaging in kidney quality assessment during normothermic machine perfusion (NMP).

**Methods** Porcine kidneys (n=22) retrieved from a local slaughterhouse underwent 2 hours of NMP. Based on creatinine clearance (CrCl) and oxygen consumption ($VO_2$), the kidneys were classified as functional (n=7) and non-functional (n=15) kidneys. PD imaging was performed at 30, 60, and 120 minutes, and PD metrics, including vascularization index (VI), flow index (FI), and vascularization flow index (VFI) were calculated. Renal blood flow (RBF), CrCl, and $VO_2$ were measured at the same time points during NMP. The metrics were compared utilizing correlation analysis.

**Results** FI and VFI moderately correlated with CrCl (r=0.537, p<0.0001; r=0.536, p<0.0001, respectively), while VI strongly correlated with $VO_2$ (r=0.839, p<0.0001). At 120 minutes, PD metrics demonstrated the highest diagnostic accuracy for distinguishing functional from non-functional kidneys, with an area under the curve (AUC) of 0.943 for VI, 0.924 for FI, and 0.943 for VFI. Cutoff values of 17% for VI, 50 a.u. for FI, and 9 a.u. for VFI provided 100% specificity and 73% sensitivity to identify non-functional kidneys, with an overall diagnostic accuracy of 82%. Baseline kidney biopsies showed moderate acute tubular necrosis in both groups with no significant differences.

**Conclusions** PD metrics strongly correlate with renal viability and effectively differentiate functional from non-functional kidneys. PD imaging can be a valuable alternative to RBF during NMP for kidney assessment.

**Introduction**

Kidney transplantation is considered the optimal treatment for end-stage renal disease.[1] Due to persistent organ shortage, kidneys from donation after circulatory death (DCD) or expanded criteria donors (ECD) are increasingly used for transplantation. Unfortunately, such kidneys are associated with a high risk of graft failure due to the suboptimal conditions of the donors such as old age, uncertain medical history, long ischemia time or pre-donation acute kidney injury.[2, 3] To reduce the risk of graft failure, organ quality has to be carefully assessed before transplantation, mainly based on procurement biopsies despite its invasive nature.[4]

Normothermic machine perfusion (NMP) has emerged as an experimental preservation technique which aims to restore and assess kidney function by supplying the organ with perfusate enriched with nutrients and oxygen at 35-37°C to simulate the physiological environment of the human body.[5, 6] This innovative platform allows for the evaluation of renal functional capacities and injury through various parameters, including renal blood flow (RBF), intrarenal resistance (IRR), urine output, creatinine clearance (CrCl), oxygen consumption ($VO_2$) and optical imaging during NMP.[7, 8]

Markgraf et al. utilized hyperspectral optical imaging (HSI) to measure renal cortex oxygen saturation and water content as indicators of tissue damage.[9, 10] Schuller et al. used magnetic resonance imaging (MRI) to assess renal blood distribution during NMP.[11, 12] Our previous research demonstrated the value of laser speckle contrast imaging (LSCI) and photoacoustic imaging (PAI) to visualize renal cortical microcirculation and assess kidney quality.[13, 14]

Alternatively, renal perfusion can be assessed using Ultrasound (US) Power Doppler (PD). Based on the Doppler effect, PD measures the amplitude of the Doppler signal and visualizes blood flow patterns.[15, 16] PD imaging has been widely used in various clinical scenarios to assess blood flow, vascular structures and hemodynamic abnormalities.[17-20] In kidney transplant patients, PD is mainly used to detect vascular complications, early acute rejection and renal dysfunction.[21-23] Despite its use in post-transplant practice, no study has reported PD imaging in assessing organ quality during pre-transplant NMP. Recognizing this gap, our study aims to investigate the additional value of PD imaging in kidney quality assessment during NMP and its correlation with function and viability markers.

## Materials and Methods

### Study design

Twenty-two kidneys from 6-month-old female Landrace pigs were harvested from a local slaughterhouse. The animals were exsanguinated before slaughtering which simulates DCD conditions. Next, kidneys were transported to the laboratory and underwent 2 hours of NMP for quality assessment.

During NMP, traditional functional parameters, including CrCl and $VO_2$, and PD data were acquired at 30, 60, and 120 minutes of NMP. The CrCl and $VO_2$ values of 120 minutes were used to classify the kidneys into functional and non-functional groups. Specifically, functional kidneys were defined as CrCl>1 ml/min/100g and $VO_2$>2.6 ml/min/100g, and non-functional kidneys were defined as CrCl⩽1 ml/min/100g and $VO_2$⩽2.6 ml/min/100g. These criteria were based on studies utilizing comparable NMP setups and perfusate components.[24, 25]

Since the kidneys were obtained from pigs intended for meat consumption, no approval from the animal ethics committee was required.

### Kidney preparation

Back-table procedures have been described elsewhere.[13] Briefly, two liters of blood were collected in a heparinized bucket and filtered by leukocyte filter (BioR 02 plus, Fresenius Kabi, the Netherlands) during exsanguination. The kidneys were then explanted by a designated butcher following a standardized procedure. Once the kidneys were explanted, back-table preparation was performed by two of the authors (Y.F. and G.A). The renal artery was cannulated and flushed with 500 mL 4°C Ringer's lactate (Baxter BV, Utrecht, the Netherlands). During transportation to the laboratory, kidneys were preserved using hypothermic machine perfusion (LifePort®, Organ Recovery Systems, USA) until NMP. Warm ischemia time (WIT) was calculated from the blood collection until cold flushing. Cold ischemia time (CIT) was defined as the hypothermic machine perfusion period.

Before NMP, the kidneys were flushed with 100 mL 4°C Ringer's lactate to remove the cold preservation solution. Subsequently, kidneys underwent 2 hours of NMP with autologous blood-based solution (Table S1) at 37°C with a controlled pressure of 70 mmHg. Carbogen (95% $O_2$ and 5% $CO_2$) was supplied via the oxygenator at flow rate of 0.5 L/min.

**Experimental setup**

The schematic of the perfusion and measurement setup is depicted in Figure 1.

The NMP system consists of a centrifugal pump head (BPX-80 Bio-Pump™ Plus, Medtronic, Minneapolis, USA) equipped with a pump drive (BVP-BP, Harvard Apparatus, Germany), an oxygenator with a heat exchanger (Hilite 1000, Medos, Germany), and a thermocirculator (E100, Lauda, Germany). A flow probe (73-4755, Harvard Apparatus, Germany) is connected in-line with the kidney and a pressure sensor (APT300, Harvard Apparatus, Germany) is directly connected to the renal artery cannula (Organ Recovery Systems, Itasca, USA). The set-up is controlled by a commercially available electronic controller (PLUGSYS Servo Controller for Perfusion, Harvard Apparatus, Germany) which enables both flow and pressure-directed perfusions.

The PD system, a Vevo3100 ultrasound (Fujifilm VisualSonics, Canada) system, is equipped with an MX201 US linear transducer array (VisualSonics, Canada) operating at 15 MHz central frequency. During the PD acquisition, the transducer was attached to a linear translational stage enabling 3D image acquisition. To provide the acoustic coupling, we covered the kidney with a polyethylene film filled with water as coupling media.

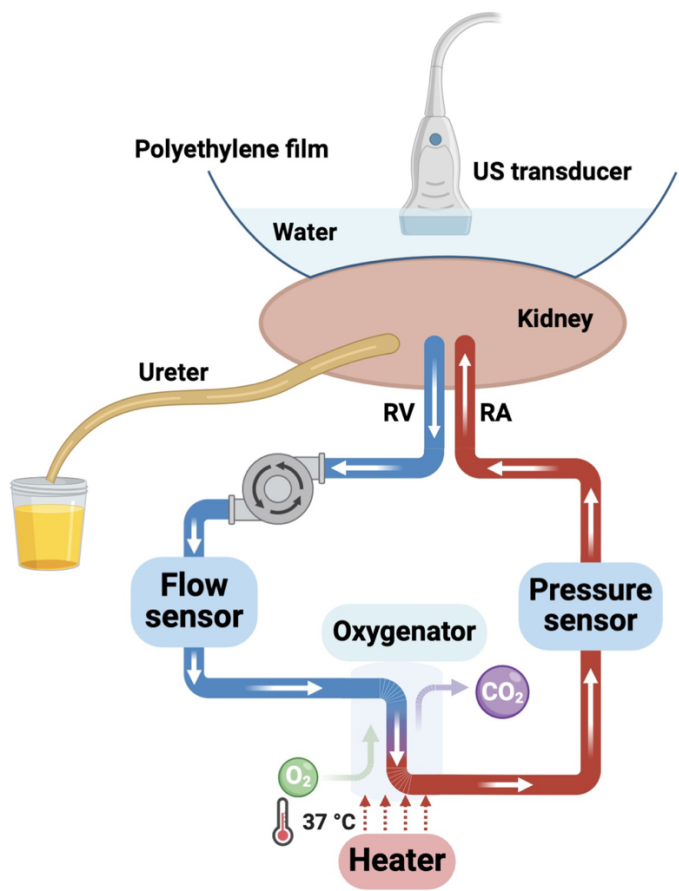

**Figure 1. Schematic of the normothermic machine perfusion (NMP) and ultrasound (US) setup. RA, renal artery; RV, renal vein.**

**Renal function, viability, and injury measurement**

RBF was monitored by the flow sensor continuously throughout NMP. The perfusate samples from the renal artery and vein, as well as urine samples, were collected at 30, 60, and 120 minutes of NMP for blood gas analysis. The urine output was recorded at the same time points.

CrCl, defined as the volume of perfusate plasma that is cleared of creatinine per unit time, is used as a renal function marker. $VO_2$, defined as the amount of oxygen consumed per unit time, is used as a renal viability marker.[26]

Baseline kidney biopsies were taken before NMP and fixed in 4% buffered paraformaldehyde. The tissue samples were then embedded in paraffin and cut into 5 μm sections. Using periodic acid-Schiff (PAS) staining, acute tubular necrosis (ATN) was evaluated and scored on a semiquantitative scale of 0 to 3 (0-no changes, 1-mild, 2-moderate, 3-severe changes) by an experienced renal pathologist (M.C.v.G.) blinded to the study. ATN score was based on the degree of brush border loss, tubular dilatation, epithelial vacuolation, thinning and sloughing, and luminal debris/casts.

**Power Doppler measurement**

PD data were acquired from the lateral aspect of the renal cortex at 30, 60, and 120 minutes of NMP. To achieve this, the kidney was positioned laterally, and the US transducer was positioned directly above the kidney. The transducer was automatically translated over a range of 45 mm by 0.5 mm step. The settings for the PD measurements are reported in Table S2, which were chosen manually to optimize the visualization of the renal cortical vasculature during NMP.

PD data were used to calculate vascularization index (VI), flow index (FI), and vascularization flow index (VFI). VI was defined as the ratio of colored pixels to all pixels within the region of interest (ROI), indicating the number of blood-filled vessels. FI was defined as the mean PD signal intensity from all colored pixels, indicating the blood volume at the time of measurement. VFI is the combination of vascularization and blood volume, mathematically multiplying VI by FI.[27] The equations are shown as below:

$$VI\ (\%) = \frac{\#\ colored\ pixels}{\#\ total\ pixels} \times 100\%, (1)$$

$$FI(a.u.) = \frac{\sum_{i=1}^{\#\ colored\ pixels} w(i)}{\#colored\ pixels}, (2)$$

$$VFI(a.u.) = VI \times FI, (3)$$

where $w(i)$ is a value of the $i^{th}$ colored pixel which represents intensity of the PD signal. The pixel value represents power of the Doppler signal and normalized based of the settings and scaled between 0 and 255. The ROI is defined as the renal cortical layer at a depth of 5 mm from the kidney surface, as the majority of functional nephrons are located within this cortical layer. For further data analysis we calculated the mean value of each metric within the acquired volume.

**Statistical analysis**

We reported continuous variables using median and interquartile range (IQR). Renal function, viability and PD metrics at each time point were compared between groups using the Mann-Whitney U test. The correlations between PD metrics and functional parameters, as well as between RBF and functional parameters were investigated using the Spearman's linear correlation test. Receiver operating characteristic (ROC) curves were used to determine the cutoff values that differentiate between functional and non-functional kidneys. Due to the organ scarcity, it is recommended to accept all potentially viable kidneys.[28, 29] Therefore, we defined cutoff values with the highest possible specificity for detecting non-functional kidneys. We adopted a significance level of 95%. We used Matlab 2023a (The MathWorks, Natick, MA, USA) and GraphPad Prism 9.3.1 (GraphPad Software Inc., San Diego, CA) for data processing, statistical analysis and data presentation.

**Results**

All kidneys underwent NMP for 2 hours without macroscopic abnormalities. Based on the pre-set criteria, 7 kidneys were classified into the functional group, and 15 kidneys were classified into the non-functional group. The average WITs were 35 minutes for the functional kidneys and 50 minutes for the non-functional kidneys, and CITs were 5.6 hours and 6.4 hours, respectively.

**Renal function and viability assessment**

RBF demonstrated significant variations between the functional and non-functional groups. Specifically, at 30, 60 and 120 minutes, RBF was 54 (45-58), 83 (66-88), and 83 (76-128) ml/min/100g in the functional group, and 39 (29-50), 49 (42-55), and 55 (43-65) ml/min/100g in the non-functional group (p=0.078, p=0.005, and p=0.003, respectively, Figure 2A). CrCl showed significant difference from 30 minutes [0.43 (0.20-1.25) vs 0.17 (0.06-0.28) ml/min/100g, p=0.039] until 120 minutes [1.61 (1.46-4.27) vs 0.33 (0.17-0.79) ml/min/100g, p<0.001, Figure 2B]. $VO_2$ diverged significantly by 60 minutes of NMP, with the functional group displaying higher $VO_2$ of 2.63 (2.34-3.03) ml$O_2$/min/100g compared to 1.82 (1.69-2.19) ml$O_2$/min/100g in the non-functional group (p=0.004, Figure 2C). Urine output was also higher in the functional group, particularly at 120 minutes [43 (28-85) vs 8 (6-18) ml/h, p=0.002, Figure 2D].

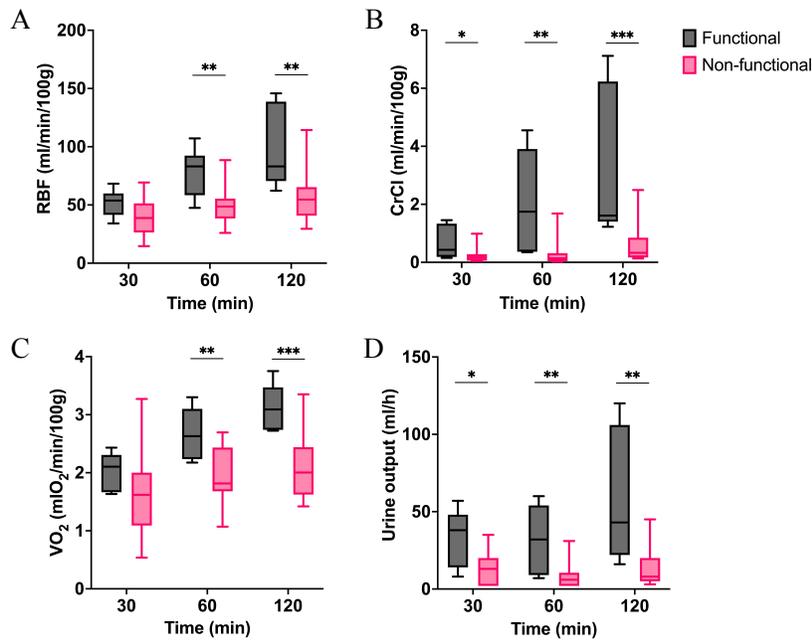

**Figure 2. Function and viability parameters measured during NMP. (A) Renal blood flow (RBF); (B) Creatinine clearance (CrCl); (C) Oxygen consumption (VO$_2$); (D) Urine output. *p≤0.05, **p≤0.01, ***p≤0.001.**

**Power Doppler metrics assessment**

The representative PD imaging frames from functional and non-functional groups are shown in Figure 3. PD images distinctly discriminate the difference in renal cortical perfusion between the groups during NMP. The functional group showed a progressive increase in signal intensity and vascularization across the time points, suggesting effective restoration of perfusion. In contrast, the non-functional group demonstrated less improvement in perfusion recovery, as indicated by lower signal intensity and vascularization percentages.

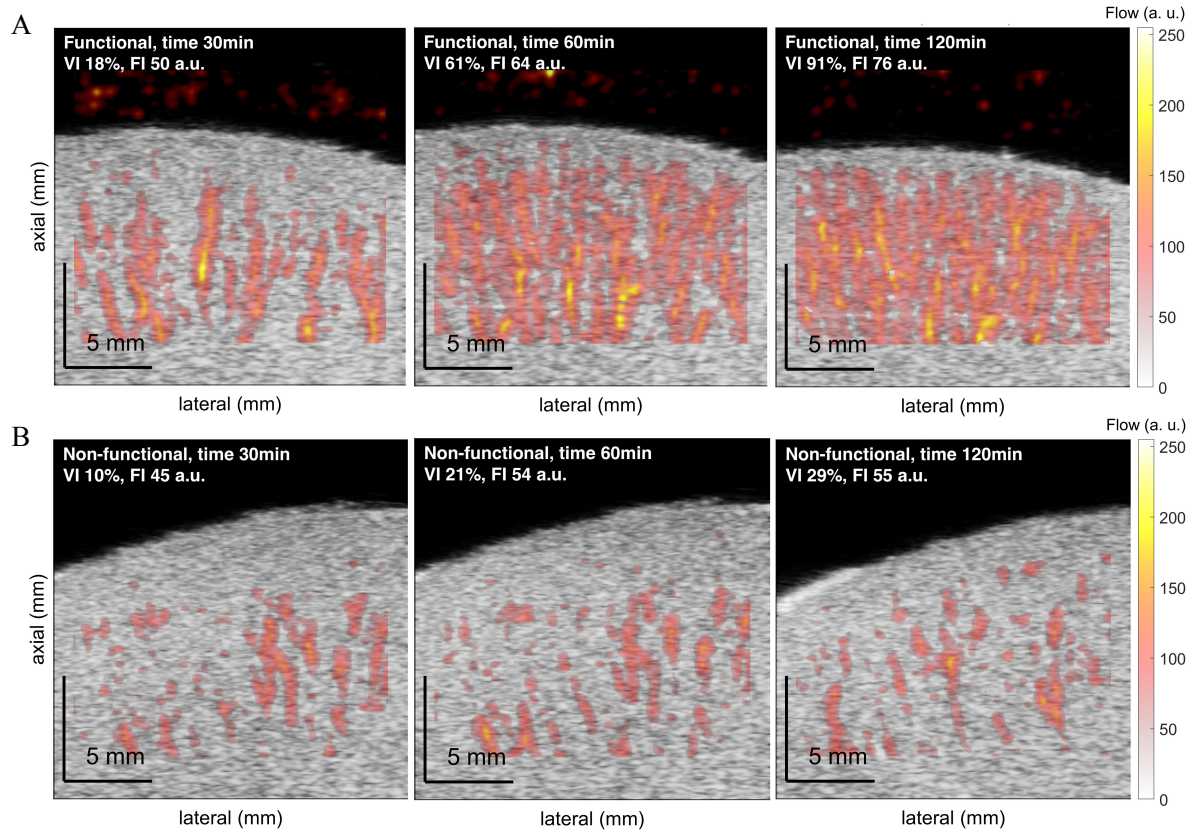

**Figure 3.** Representative Power Doppler (PD) imaging frames of renal cortex in (A) Functional group and (B) Non-functional group at 30, 60, and 120 minutes of NMP. FI, flow index; VI, vascularization index.

The values of VI, FI, and VFI are depicted in Figure 4. VI varied significantly between the functional and non-functional groups at 30 minutes of NMP [9 (7-17)% vs 2 (1-8)%, p=0.026, Figure 4A]. FI also differed between the two groups. The functional group showed a significantly higher FI at 30 minutes, indicating more robust perfusion compared to the non-functional group [46 (44-51) vs 36 (33-44) a.u., p=0.032, Figure 4B]. VFI showed similar trends to VI and FI. At 30 minutes, the functional group demonstrated significantly higher VFI compared to the non-functional group [5 (4-10) vs 1 (0-4) a.u., p=0.026, Figure 4C].

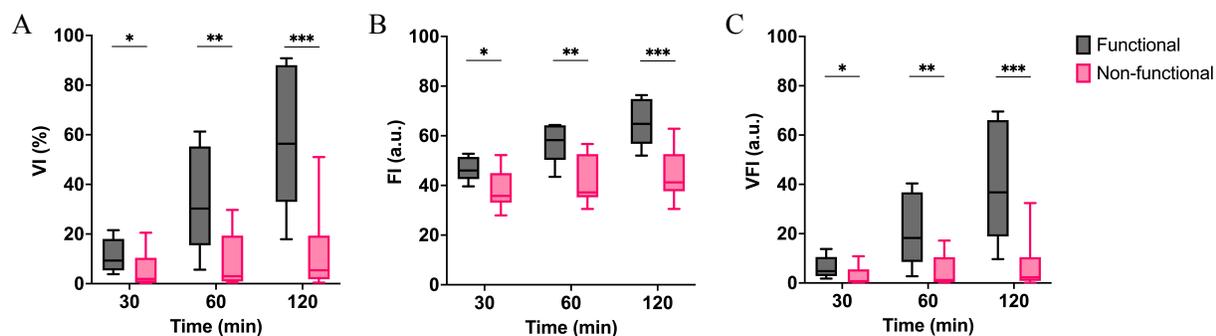

**Figure 4. Quantitative PD metrics showing significant differences between the functional and non-functional groups at 30 minutes of NMP.** (A) Vascularization index (VI); (B) Flow index (FI); (C) Vascularization flow index (VFI). *p≤0.05, **p≤0.01, ***p≤0.001.

**Correlation between RBF and functional metrics**

We investigated the correlation between renal perfusion velocities and functional parameters as depicted in Figure 5. Both RBF and PD metrics demonstrated significant positive correlations with CrCl and VO2. Specifically, while the correlation between RBF and CrCl was low positive (r=0.423, p<0.001), PD metrics demonstrated moderate correlations with CrCl (VI, r=0.528, p<0.0001; FI, r=0.537, p<0.0001; VFI, r=0.536, p<0.0001). Notably, both RBF and PD metrics showed strong positive correlations with VO$_2$, while VI exhibited the highest correlation coefficient (r=0.839, p<0.0001).

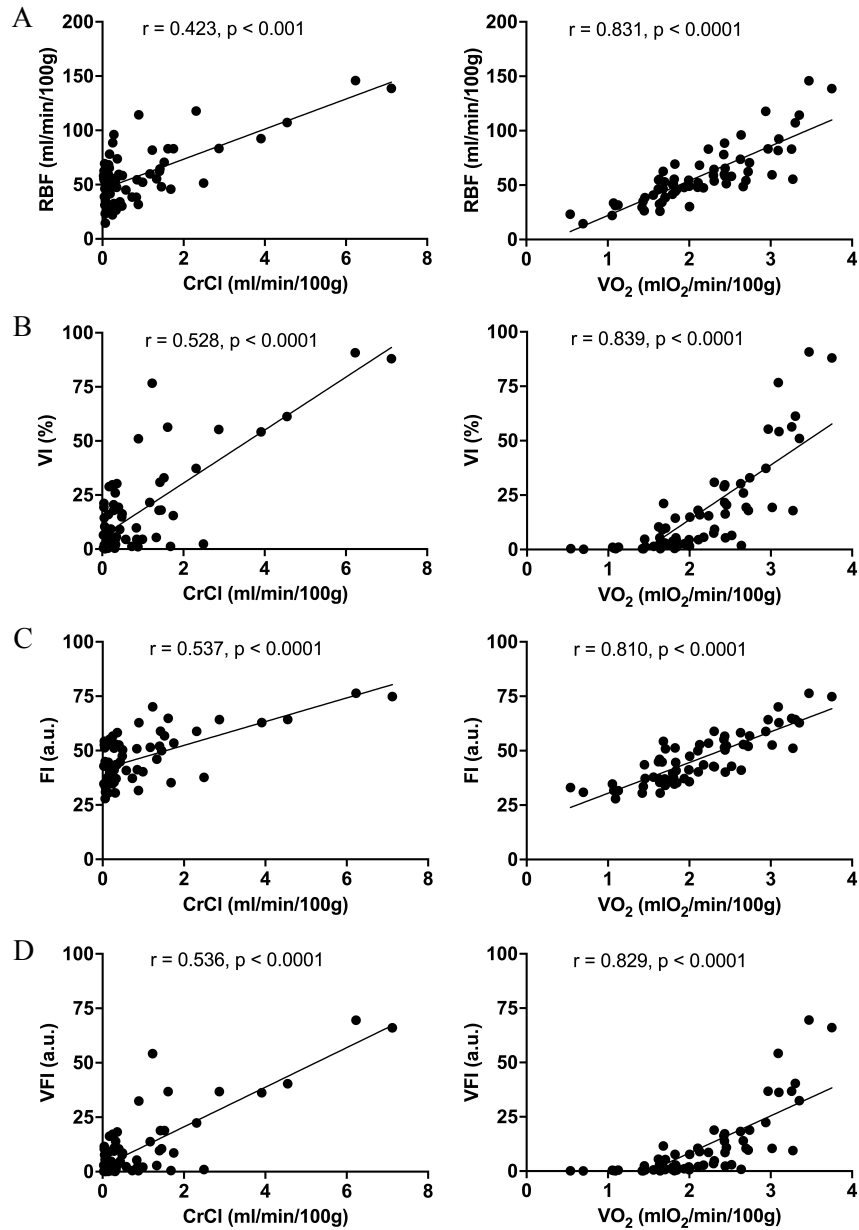

**Figure 5.** Correlation between (A) RBF, (B) VI, (C) FI, (D) VFI and CrCl (left) and $VO_2$ (right). Black scatters represent all kidneys at all measured time points.

**Cutoff values for intergroup differentiation**

The ROC curve analyses of PD metrics and RBF at 30, 60, and 120 minutes of NMP are summarized(shown?) in Figure S1, S2 and Table S3, S4. Among these time points, PD metrics at 120 minutes showed the most robust cutoff values for differentiating functional and non-functional kidneys, with an area under the curve (AUC) of 0.943 for VI, 0.924 for FI, and 0.943 for VFI. Similarly, RBF at 120 minutes showed the highest AUC of 0.886 (p=0.004).

To ensure that no functional kidney is rejected, we analyzed the cutoff values for these metrics to achieve 100% specificity in discriminating non-functional kidneys. Compared to RBF which shows a negative predictive value (NPV) of 58% and overall accuracy of 77%, cutoff values of 17% for VI, 50 a.u. for FI, and 9 a.u. for VFI demonstrated higher NPVs of 64% and overall diagnostic accuracy of 82%.

**Renal injury assessment**

Representative histological images of kidney biopsies are depicted in Figure 6. In both groups, the tubules exhibited loss of brush border, debris, and lumen dilation, suggesting moderate ATN prior to NMP. No significant difference of the semiquantitative scale was observed between the groups.

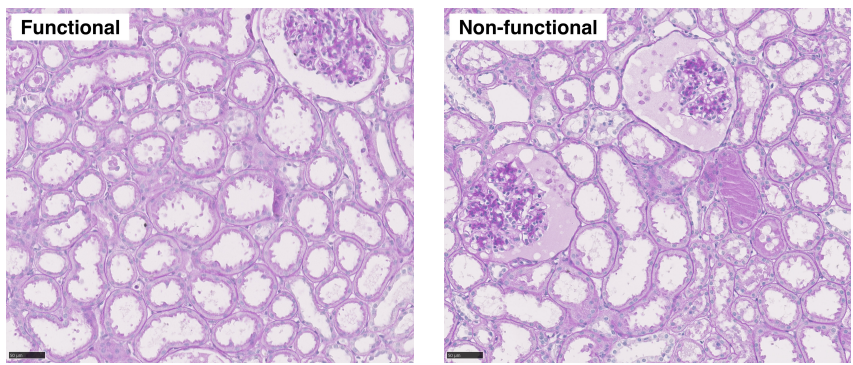

**Figure 6. Representative periodic-acid-Schiff (PAS) staining of the functional group (left), and non-functional group (right) prior to NMP. 40 × magnification.**

**Discussion**

This study used a porcine kidney NMP model to investigate the value of PD imaging in assessing kidney quality. We observed significant differences in PD metrics between functional and non-functional kidneys. Our findings revealed a moderate positive correlation between PD metrics and renal function, and a strong positive correlation between PD metrics and renal viability. Compared to RBF, PD imaging may provide earlier detection of non-functional kidneys in this porcine model. PD imaging has unique advantages in ease of use, real-time monitoring, and relatively low cost. Since NMP is gradually being introduced into clinical practice in academic medical centers,[30, 31] it is feasible to apply PD imaging for pre-transplant kidney quality assessment in clinical settings.

In functional kidneys, PD imaging displays a diffused, homogeneous blush color across the entire renal cortex. In contrast, non-functional kidneys significantly lower VI, FI, and VFI, indicating fewer blood-filled vessels and reduced blood volume in the renal cortex. It can be explained by the prolonged WIT and high sensitivity to ischemic injury that result in high intrarenal resistance and inferior blood flow during machine perfusion.[32-34]

Compared to RBF as routine monitoring during NMP, PD metrics differentiated between functional and non-functional kidneys earlier. This discrepancy reflects the heterogeneity of the perfusion in the entire kidney (RBF) versus renal cortical perfusion status (PD), which is in line with the study by Schutter et al.[11] which reported heterogeneous corticomedullar ratios in kidneys with consistent RBF during NMP.

Post-transplant PD imaging assessments can be affected by extrarenal variables rather than inherent renal injury. Renal cortical blood perfusion is easily affected by blood pressure, renal capillary wedge pressure, and vascular compliance, which reflects systemic hemodynamics.[35] In contrast, pre-transplant assessment with standardized perfusion setups can reduce bias from extrarenal variables. We investigated the correlation between PD metrics and function and viability markers. VI, FI, and VFI showed a better correlation with CrCl than RBF. This can be explained by the fact that the majority of functional nephrons are located in the renal cortex. However, due to the complex mechanisms of kidney function, predicting renal function solely based on PD imaging is not ideal. These metrics need to be combined with other characteristics, biomarkers and biophysical properties to accurately evaluate function. In terms of renal viability, both RBF and PD metrics showed a strong correlation with $VO_2$, with VI displaying the highest correlation coefficient, indicating that PD metrics are reliable markers of renal metabolic activity.

Our results showed that, with appropriate cutoff values, both PD metrics and RBF can achieve 100% specificity in identifying non-functional kidneys, with PD showing higher overall accuracy. Based on this, kidneys classified as non-functional are strongly recommended to discard. However, it is important to note that with an overall accuracy of 82%, 18% of kidneys potentially fall into the "false negative" category. This underscores the challenge of relying solely on one factor or technique to predict kidney transplantability. Incorporating PD imaging into the current scoring system as an alternative to RBF might enhance its accuracy.[8, 36]

Interestingly, histological results revealed no significant differences between the groups. One explanation could be sampling error: the local evaluation of the renal parenchyma cannot represent the overall condition of the kidney.[37]

The present study has several limitations to be addressed. Firstly, the use of slaughterhouse kidneys may exhibit higher interindividual variability in kidney viability and extent of renal injury, which could have interfered with our outcome measurements. Secondly, while porcine kidneys are commonly utilized in preclinical machine perfusion experiments, their anatomical and physiological characteristics differ from human kidneys. It is important to exercise caution when extrapolating our findings to clinical scenarios. Moreover, we did not transplant the kidneys and hence not able to measure the posttransplant renal function. Further research should focus on applying PD imaging to experimental kidneys intended for transplantation to validate our findings.

## Conclusion

This study demonstrated the potential of PD imaging in pre-transplant kidney quality assessment using a porcine kidney NMP model. The PD metrics strongly correlate with renal viability and can effectively identify non-functional kidneys. PD imaging can be a valuable alternative to RBF during NMP. The explorative translation into clinical practice might help to optimize the utilization of marginal kidneys.

## Acknowledgement

Imaging experiments were performed on the LAZR-X system (donation Josephine Nefkens Personalized Cancer treatment Program) in collaboration with the Applied Molecular Imaging core facility (AMIE) of the Erasmus MC. The authors also thank Slagerij van Meurs for providing porcine kidneys for research, and Triallab of the Erasmus MC for the help on perfusate analyses.

## Conflict of Interest

Gijs van Soest was the PI on research projects, conducted at Erasmus MC, that received research support from FUJIFILM VisualSonics, Shenzhen Vivolight, Boston Scientific, Waters and Mindray.

## Funding

The work was supported by the "HOST" project funded by the Dutch Research Council (NWO, project number 20719).

## Author contributions

YF participated in the research design, experiments, data analysis and interpretation, and drafting the manuscript. AN participated in the research design, experiments, data analysis and interpretation, and reviewing the manuscript. JE, GA and MCvG participated in the experiments, data analysis and interpretation, and reviewing the manuscript. RM, RdB, and GvS participated in the research design and critically reviewing the manuscript.

## References

1. Tonelli M, Wiebe N, Knoll G, Bello A, Browne S, Jadhav D, et al. Systematic review: kidney transplantation compared with dialysis in clinically relevant outcomes. Am J Transplant. 2011;11(10):2093-109.
2. Port FK, Bragg-Gresham JL, Metzger RA, Dykstra DM, Gillespie BW, Young EW, et al. Donor characteristics associated with reduced graft survival: an approach to expanding the pool of kidney donors. Transplantation. 2002;74(9):1281-6.
3. Rao PS, Ojo A. The alphabet soup of kidney transplantation: SCD, DCD, ECD--fundamentals for the practicing nephrologist. Clin J Am Soc Nephrol. 2009;4(11):1827-31.
4. von Moos S, Akalin E, Mas V, Mueller TF. Assessment of Organ Quality in Kidney Transplantation by Molecular Analysis and Why It May Not Have Been Achieved, Yet. Front Immunol. 2020;11:833.
5. Mazilescu LI, Urbanellis P, Kim SJ, Goto T, Noguchi Y, Konvalinka A, et al. Normothermic Ex Vivo Kidney Perfusion for Human Kidney Transplantation: First North American Results. Transplantation. 2022;106(9):1852-9.
6. Hosgood SA, Callaghan CJ, Wilson CH, Smith L, Mullings J, Mehew J, et al. Normothermic machine perfusion versus static cold storage in donation after circulatory death kidney transplantation: a randomized controlled trial. Nature Medicine. 2023;29(6):1511-9.
7. Hamelink TL, Ogurlu B, De Beule J, Lantinga VA, Pool MBF, Venema LH, et al. Renal Normothermic Machine Perfusion: The Road Toward Clinical Implementation of a Promising Pretransplant Organ Assessment Tool. Transplantation. 2022;106(2):268-79.
8. Hosgood SA, Barlow AD, Hunter JP, Nicholson ML. Ex vivo normothermic perfusion for quality assessment of marginal donor kidney transplants. Br J Surg. 2015;102(11):1433-40.
9. Markgraf W, Feistel P, Thiele C, Malberg H. Algorithms for mapping kidney tissue oxygenation during normothermic machine perfusion using hyperspectral imaging. Biomed Tech (Berl). 2018;63(5):557-66.
10. Markgraf W, Lilienthal J, Feistel P, Thiele C, Malberg H. Algorithm for Mapping Kidney Tissue Water Content during Normothermic Machine Perfusion Using Hyperspectral Imaging. Algorithms. 2020;13(11):289.

11. Schutter R, Lantinga VA, Hamelink TL, Pool MBF, van Varsseveld OC, Potze JH, et al. Magnetic resonance imaging assessment of renal flow distribution patterns during ex vivo normothermic machine perfusion in porcine and human kidneys. Transpl Int. 2021;34(9):1643-55.
12. Hamelink TL, Ogurlu B, Pamplona CC, Castelein J, Bennedsgaard SS, Qi H, et al. Magnetic resonance imaging as a noninvasive adjunct to conventional assessment of functional differences between kidneys in vivo and during ex vivo normothermic machine perfusion. Am J Transplant. 2024.
13. Fang YT, van Ooijen L, Ambagtsheer G, Nikolaev AV, Clahsen-van Groningen MCCV, Dankelman J, et al. Real-time laser speckle contrast imaging measurement during normothermic machine perfusion in pretransplant kidney assessment. Laser Surg Med. 2023;55(8):784-93.
14. Nikolaev AV, Fang Y, Essers J, Panth KM, Ambagtsheer G, Clahsen-van Groningen MC, et al. Pre-transplant kidney quality evaluation using photoacoustic imaging during normothermic machine perfusion. Photoacoustics. 2024;36:100596.
15. Martinoli C, Pretolesi F, Crespi G, Bianchi S, Gandolfo N, Valle M, et al. Power Doppler sonography: clinical applications. European Journal of Radiology. 1998;27:S133-S40.
16. Szabo TL. Chapter 11 - Doppler Modes. In: Szabo TL, editor. Diagnostic Ultrasound Imaging: Inside Out (Second Edition). Boston: Academic Press; 2014. p. 431-500.
17. Aziz MU, Eisenbrey JR, Deganello A, Zahid M, Sharbidre K, Sidhu P, et al. Microvascular Flow Imaging: A State-of-the-Art Review of Clinical Use and Promise. Radiology. 2022;305(2):250-64.
18. Smith E, Azzopardi C, Thaker S, Botchu R, Gupta H. Power Doppler in musculoskeletal ultrasound: uses, pitfalls and principles to overcome its shortcomings. J Ultrasound. 2021;24(2):151-6.
19. Bhasin S, Cheung PP. The Role of Power Doppler Ultrasonography as Disease Activity Marker in Rheumatoid Arthritis. Dis Markers. 2015;2015.
20. Yamasato K, Zalud I. Three dimensional power Doppler of the placenta and its clinical applications. J Perinat Med. 2017;45(6):693-700.
21. Franke D. The diagnostic value of Doppler ultrasonography after pediatric kidney transplantation. Pediatr Nephrol. 2022;37(7):1511-22.
22. Shebel HM, Akl A, Dawood A, El-Diasty TA, Shokeir AA, Ghoneim MA. Power Doppler Sonography in Early Renal Transplantation: Does It Differentiate Acute Graft Rejection from Acute Tubular Necrosis? Saudi J Kidney Dis T. 2014;25(4):733-40.
23. Datta R, Sandhu M, Saxena AK, Sud K, Minz M, Suri S. Role of duplex Doppler and power Doppler sonography in transplanted kidneys with acute renal parenchymal dysfunction. Australas Radiol. 2005;49(1):15-20.
24. Ogurlu B, Pamplona CC, Van Tricht IM, Hamelink TL, Lantinga VA, Leuvenink HGD, et al. Prolonged Controlled Oxygenated Rewarming Improves Immediate Tubular Function and Energetic Recovery of Porcine Kidneys During Normothermic Machine Perfusion. Transplantation. 2023;107(3):639-47.
25. Markgraf W, Malberg H. Preoperative Function Assessment of Ex Vivo Kidneys with Supervised Machine Learning Based on Blood and Urine Markers Measured during Normothermic Machine Perfusion. Biomedicines. 2022;10(12).
26. Venema LH, van Leeuwen LL, Posma RA, van Goor H, Ploeg RJ, Hannaert P, et al. Impact of Red Blood Cells on Function and Metabolism of Porcine Deceased Donor Kidneys During Normothermic Machine Perfusion. Transplantation. 2022;106(6):1170-9.
27. Odibo AO, Goetzinger KR, Huster KM, Christiansen JK, Odibo L, Tuuli MG. Placental volume and vascular flow assessed by 3D power Doppler and adverse pregnancy outcomes. Placenta. 2011;32(3):230-4.

28. Hobeika MJ, Miller CM, Pruett TL, Gifford KA, Locke JE, Cameron AM, et al. PROviding Better ACcess To ORgans: A comprehensive overview of organ-access initiatives from the ASTS PROACTOR Task Force. Am J Transplant. 2017;17(10):2546-58.
29. Cooper M, Formica R, Friedewald J, Hirose R, O'Connor K, Mohan S, et al. Report of National Kidney Foundation Consensus Conference to Decrease Kidney Discards. Clin Transplant. 2019;33(1):e13419.
30. Hosgood SA, Callaghan CJ, Wilson CH, Smith L, Mullings J, Mehew J, et al. Normothermic machine perfusion versus static cold storage in donation after circulatory death kidney transplantation: a randomized controlled trial. Nat Med. 2023;29(6):1511-9.
31. Rijkse E, de Jonge J, Kimenai H, Hoogduijn MJ, de Bruin RWF, van den Hoogen MWF, et al. Safety and feasibility of 2 h of normothermic machine perfusion of donor kidneys in the Eurotransplant Senior Program. BJS Open. 2021;5(1).
32. Birk AV, Liu S, Soong Y, Mills W, Singh P, Warren JD, et al. The mitochondrial-targeted compound SS-31 re-energizes ischemic mitochondria by interacting with cardiolipin. J Am Soc Nephrol. 2013;24(8):1250-61.
33. Kaasik A, Safiulina D, Zharkovsky A, Veksler V. Regulation of mitochondrial matrix volume. Am J Physiol Cell Physiol. 2007;292(1):C157-63.
34. Liu S, Soong Y, Seshan SV, Szeto HH. Novel cardiolipin therapeutic protects endothelial mitochondria during renal ischemia and mitigates microvascular rarefaction, inflammation, and fibrosis. Am J Physiol Renal Physiol. 2014;306(9):F970-80.
35. O'Neill WC. Renal resistive index: a case of mistaken identity. Hypertension. 2014;64(5):915-7.
36. Hosgood SA, Thompson E, Moore T, Wilson CH, Nicholson ML. Normothermic machine perfusion for the assessment and transplantation of declined human kidneys from donation after circulatory death donors. Br J Surg. 2018;105(4):388-94.
37. Dhaun N, Bellamy CO, Cattran DC, Kluth DC. Utility of renal biopsy in the clinical management of renal disease. Kidney Int. 2014;85(5):1039-48.

**Supplementary**

**Table S1. Components of the normothermic machine perfusion perfusate**

| Priming | Volume |
|---|---|
| Ringer's lactate (Baxter, the Netherlands) | 300 |
| Autologous leukocyte-depleted blood | 500 |
| **Additives** | |
| 8.4% sodium bicarbonate (B. Braun, Germany) | 10 |
| 5% glucose (Baxter, the Netherlands) | 10 |
| Verapamil (Sigma-Aldrich, the Netherlands) | 1 |
| Dexamethasone (Centrafarm, the Netherlands) | 0.3 |
| 90mg creatinine (Sigma-Aldrich, the Netherlands) | - |
| **Infusion 20mL/h** | |
| Aminoplasmal (B. Braun, Germany) | 45 |
| 8.4% sodium bicarbonate (B. Braun, Germany) | 1.5 |
| 100 U/mL insulin (NovoRapid®, Denmark) | 0.5 |

**Table S2. Settings used for PD data acquisition**

| PRF (kHz) | Tx frequency (MHz) | Gate | Wall filter | Sensitivity | Dynamic range (dB) |
|---|---|---|---|---|---|
| 2 | 12.5 | 4 | High | 5 | 30 |

**Table S3. ROC curve analysis of PD metrics**

| Time | PD metrics | AUC | 95% CI | p value |
|---|---|---|---|---|
| 30 min | VI (%) | 0.800 | 0.616-0.984 | 0.026 |
|  | FI (a.u.) | 0.791 | 0.603-0.978 | 0.032 |
|  | VFI (a.u.) | 0.800 | 0.616-0.984 | 0.026 |
| 60 min | VI (%) | 0.867 | 0.710-1.000 | 0.007 |
|  | FI (a.u.) | 0.886 | 0.742-1.000 | 0.004 |
|  | VFI (a.u.) | 0.867 | 0.710-1.000 | 0.007 |
| 120 min | VI (%) | 0.943 | 0.848-1.000 | 0.001 |
|  | FI (a.u.) | 0.924 | 0.814-1.000 | 0.002 |
|  | VFI (a.u.) | 0.943 | 0.848-1.000 | 0.001 |

AUC, area under the curve; 95% CI, 95% confidence interval; FI, flow index; PD, Power Doppler; VFI, vascularization flow index; VI, vascularization index.

**Table S4. ROC curve analysis of RBF**

| Time | AUC | 95% CI | p value |
|---|---|---|---|
| 30 min | 0.748 | 0.535-0.960 | 0.067 |
| 60 min | 0.867 | 0.692-1.000 | 0.007 |
| 120 min | 0.886 | 0.747-1.000 | 0.004 |

AUC, area under the curve; 95% CI, 95% confidence interval; RBF, renal blood flow.

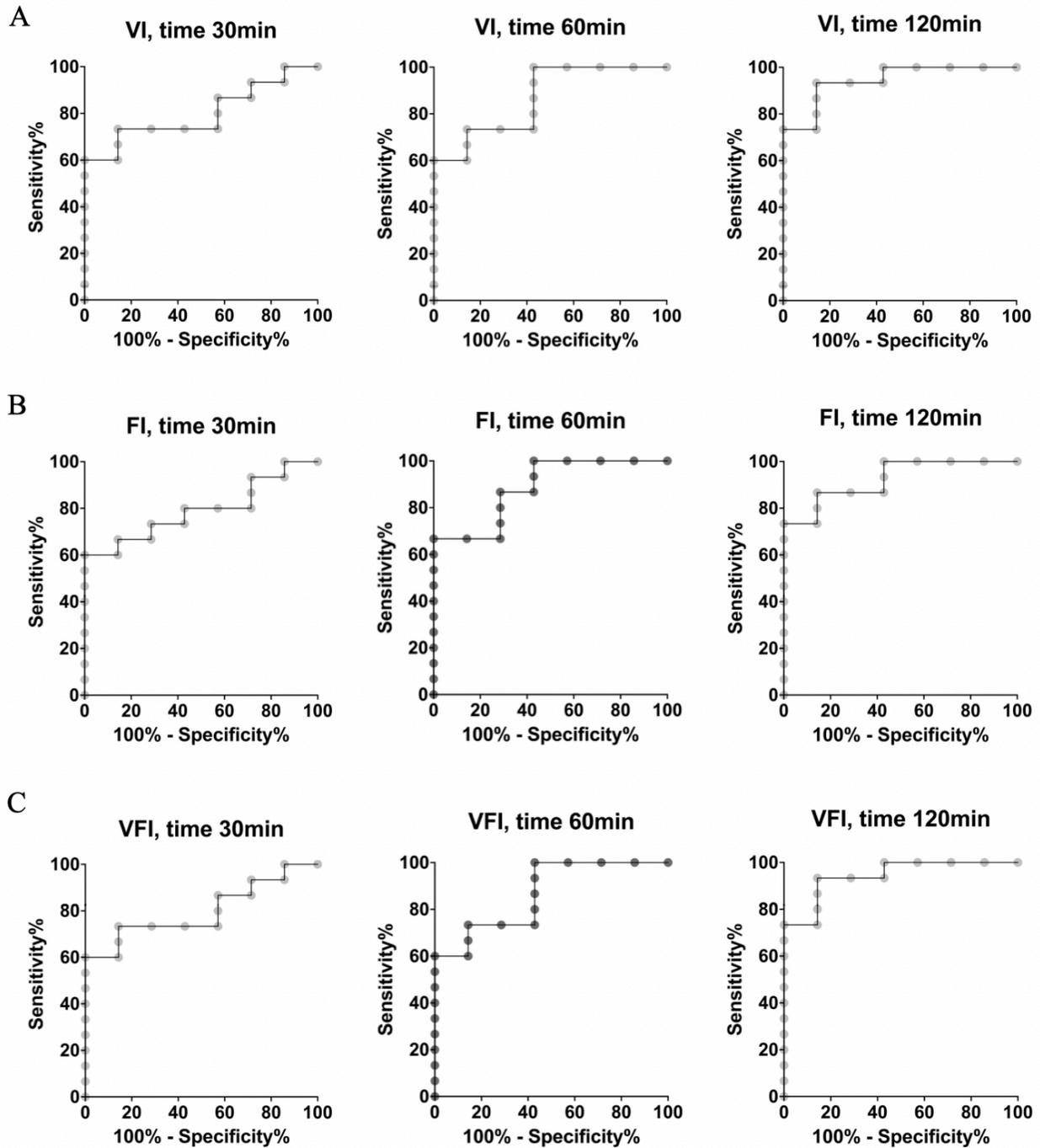

**Figure S1.** Receiver operating characteristic (ROC) curves of (A) vascularization index (VI), (B) flow index (FI), and (C) vascularization flow index (VFI) to differentiate between the functional and non-functional kidneys at 30, 60, 120 minutes of NMP.

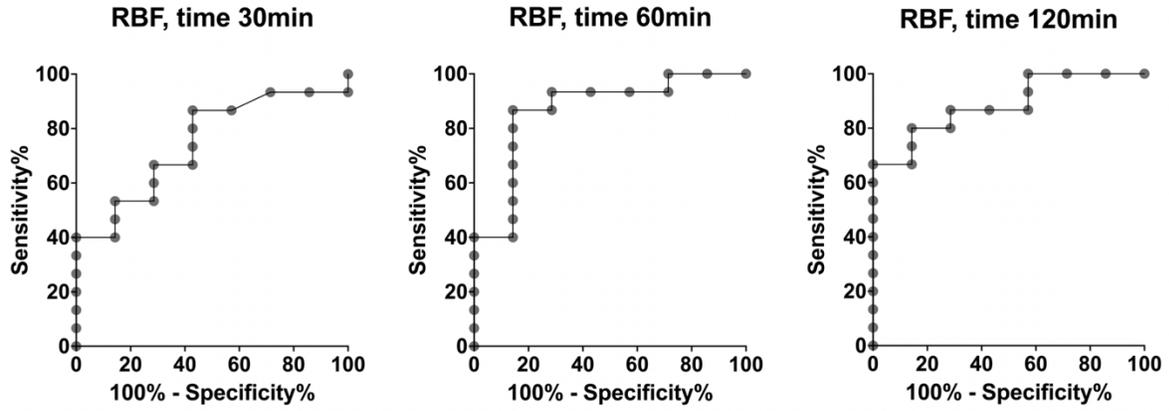

**Figure S2.** Receiver operating characteristic (ROC) curves of renal blood flow (RBF) to differentiate between the functional and non-functional kidneys at 30, 60 and 120 minutes of NMP.